\documentclass[a4,10pt,twoside]{article}

\usepackage{pst-all}
\usepackage{pstricks}
\usepackage{psfrag}
\usepackage[dvips]{graphicx}
\usepackage{graphics}
\usepackage{amsmath,amsthm,amssymb}
\usepackage[small,sc]{caption2}
\usepackage{multicol}
\usepackage{anysize}
\marginsize{2.5cm}{2.5cm}{2.cm}{2.cm} 

\newcommand{\Fp}{\ensuremath{\mathcal{F^{\,\prime}}}}

\newcommand{\A}{\ensuremath{\mathcal{A}}}
\newcommand{\F}{\ensuremath{\mathcal{F}}}
\newcommand{\Ap}{\ensuremath{\mathcal{A^{\,\prime}}}}
\newcommand{\Bp}{\ensuremath{\mathcal{B^{\,\prime}}}}

\newcommand{\Mn}[1]{\ensuremath{\text{Mn}^{#1+}}}
\newcommand{\ZB}{1$^{\rm ra}$ZB }

\newcommand{\comp}[2]{#1$_{1-x}$#2$_x$MnO$_3$}
\newcommand{\Com}[4]{#1$_{#2}$#3$_{#4}$MnO$_3$}

\newcommand{\Coml}[4]{#1$_{#2}$#3$_{#4}$MnO$_4$}

\newpsobject{grilla}{psgrid}{subgriddiv=1,griddots=5,gridlabels=6pt}  

\definecolor{navy}{rgb}{0,0,128}

\title{Spin Excitations in Half-Doped Manganites}
\author{ Ivon R. Buitrago $^1$  and  C.I. Ventura$^2$ \thanks{e-mail: ventura@cab.cnea.gov.ar}\\
  \small $^1$ Instituto Balseiro, Univ. Nac. de Cuyo, BarilochInstituto Balseiro, Univ. Nac. de Cuyo, Barilochee, \\
  \small $^2$ Centro At\'omico Bariloche and Univ. Nac. de R{\'{\i}}o Negro, Bariloche, ARGENTINA.}

\date{}

\begin{document}
\maketitle
\begin{abstract}
  We investigate magnetic excitations in half-doped colossal magnetoresistance manganites. In particular, we focus on spin excitations in the CE phase originally proposed by Goodenough (Phys. Rev. 100, 564 (1955)).  Using a localized spin model we calculated magnons for 3D-perovskite compounds such as La$_{1-x}$M$_x$MnO$_3$, where M=Ca,Sr,Ba, and for their 2D-laminar counterparts. We compared them with predictions for the spin excitations corresponding to other phases proposed. For the laminar half-doped manganite La$_{0.5}$Sr$_{1.5}$MnO$_4$, for which  magnon measurements by inelastic neutron scattering exist, as well as an estimation of the magnetic couplings, our calculations  agree well  with the experimental data.  \\
  
\end{abstract}

\begin{multicols}{2}

  \subsection*{Introduction}
  Colossal magnetoresistance (CMR) manganites have been of great interest to the scientific community for several decades because of their potential for applications in magnetic and electronic devices due to their rich phase diagrams as a function of concentration. In 1955, Goodenough~\cite{Goodenough} proposed the antiferromagnetic "CE" type of magnetic, charge  and orbital  ordering below the Neel temperature for manganite perovskite oxides at half-doping (\comp{La}{M}, at $x=0.5$). It consists of planes with   ferromagnetic (FM)  zig-zag chains of  alternating $\Mn{3}$ and $\Mn{4}$ spins, coupled antiferromagnetically (AF) between them. Between planes, a difference arises between the magnetic and charge orderings in the CE phase:  the spins form a checkerboard arrangement, while equal charges are stacked  one on top of each other. However, with the neutron diffraction study of \Com{Pr}{0.6}{Ca}{0.4} by Daoud-Aladine \emph{et al.}~\cite{Aladine}   in 2002, the realization of the CE  ground state for these compounds was questioned, and the Zener polarons or "dimers" phase was proposed. In the latter, an electron is considered to be delocalized between two nearest-neighbour (NN)  Mn sites of a zig-zag chain, such that they can be thought to form ferromagnetic dimers with an effective intermediate valence for all Mn ions ($\Mn{3.5}$).  
  
  As the presence of  twin boundaries in half-doped perovkite manganites hindered magnon measurements with inelastic neutron scattering (INS), except at very low energies near the center of the \ZB, in 2006 Senff \emph{et al.}~\cite{Senff, Braden}  undertook INS experiments for a 2D-laminar half-doped manganite: \Coml{La}{0.5}{Sr}{1.5}. They succeeded in measuring well defined magnons  and, from their  fit  of  the magnon dispersion relations in terms of a two-dimensional  CE phase  with $\Mn{3}$ and $\Mn{4}$ localized moments, concluded that only a CE  ground state could explain the spin excitation data.  
  
  Due to the still open discussion concerning the two main possible scenarios for the ground state of half-doped manganites, we decided to calculate the magnetic excitations of the CE phase in layered and in perovskite manganites, and compare them with those characteristic of other competing phases.  
  
  \subsection*{Calculation of magnon spin excitations} 
  Our approach is based on previous work by Ventura and Alascio~\cite{Ventura} of 2003. Using a model of localized magnetic moments, magnons for the half-doped perovskite manganites were calculated~\cite{Ventura} for the dimer phase, and for a
  charge ordered phase (CO): having the same spin arrangement of the CE-phase and a slight difference in the charge order, i.e. between planes, alternate $\Mn{3}$ and $\Mn{4}$ charges are stacked on top of each other.\\  
  \begin{center}  
    \scalebox{0.5}{
      \begin{pspicture}(0,0)(12,18) 
        \psfrag{x}{\Large{$x$}}
        \psfrag{y}{\Large{$y$}}
        \psfrag{z}{\Large{$z$}}
        \psfrag{Sx}{\Large{$s_x$}}
        \psfrag{Sy}{\Large$s_y$}
        \psfrag{Sz}{\Large$s_z$}
        \psfrag{a}{\Large$a_0$}
        \psfrag{c}{\Large$c$}
        \psfrag{C1}{\Large{I}}
        \psfrag{C2}{\Large{II}}
        \psfrag{C3}{\Large{III}}
        \rput[c]{U}(6,9){\includegraphics[width=0.7\textwidth]{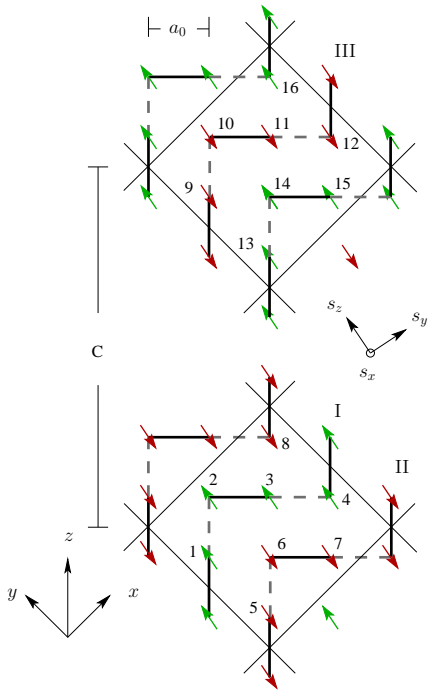}}
    \end{pspicture}}
    \captionof{figure}{Schematic representation of the CE-3D phase. $\F$: FM inter-dimer (dashed line) and $\Fp$: FM intra-dimer (solid line) couplings along a zig-zag chain. $\A$: AF intra-plane coupling, between zig-zag chains. $\Ap \big({\scriptsize\Mn{3}-\Mn{3}}\big)$ and $\Bp \big({\scriptsize\Mn{4}-\Mn{4}}\big)$: the AF inter-plane couplings, depending on the Mn-charge.} 
    \label{fig1}
\end{center}

 To determine the spin excitations of the CE-phase, we consider the three-dimensional unit cell with 16 spins $[$with $S_1=2$ ($\Mn{3}$) and $S_2=1.5$ (\Mn{4})$]$ schematically represented in Figure~\ref{fig1}, In order to be able to discuss other phases of interest, such as the dimer phase, we consider two ferromagnetic NN coupling parameters along the zig-zag chains: $\F$ and $\Fp$, and an antiferromagnetic coupling $\A$ between chains in the same plane. With respect to the AF coupling between planes, we take into account that in the CE phase each $\Mn{3}$ ion in a given plane has as NN in adjacent planes other $\Mn{3}$ ions, while the same respectively happens for the $\Mn{4}$ ions. Therefore, our new ingredient for the description of the CE phase is the inclusion of two AF inter-plane coupling parameters (between adjacent planes),
 depending on the charge considered: $\Ap \big({\scriptsize\Mn{3}-\Mn{3}}\big)$ and $\Bp \big({\scriptsize\Mn{4}-\Mn{4}}\big)$ instead of just one AF inter-plane coupling parameter as in the previous CO-phase calculation~\cite{Ventura}. In the latter, independent of the charge, all NN Mn-pairs between adjacent planes are alike: in the
 sense that they involve one $\Mn{3}$ and one $\Mn{4}$ ion, thus justifying the introduction of a single AF-interplane coupling. 

 In our simplified model we can describe the CE, dimer and CO phases by considering the spin values and magnetic couplings as in Table~\ref{tab.1}.
 \begin{center}
   \begin{tabular}{|c|c|c|}
     \hline
     \text{CE phase} & \text{Dimer phase} & \text{CO phase} \\\hline\hline
     $S_1\ne=S_2$ & $S_1=S_2$ & $S_1\ne S_2$ \\\hline
     $F=\Fp$ & $\F\ne \Fp$ & $F=\Fp$ \\\hline
     $\Ap\ne\Bp$ & $\Ap=\Bp=$ & $\Ap=\Bp$ \\\hline
   \end{tabular}
   \captionof{table}{Description of the CE, dimer and CO phases in terms of our model parameters.} 
   \label{tab.1}
 \end{center}

 Taking into account all NN couplings between Mn-spins, the Heisenberg Hamiltonian for the CE-3D phase is given by: 
 
 \begin{align}
   & H=-\Fp\,\sum_{\langle n,n'\rangle\in C,D}\,S_n\cdot S_{n'} -  \F\,\sum_{\langle n,n'\rangle\in C,\notin\,D}\,S_n\cdot S_{n'}\notag\\ 
   &+\A\,\sum_{\langle n,n'\rangle/\in\,P,\notin C}\,S_n\cdot S_{n'} +\Ap\,\sum_{\langle n,n'\rangle/\notin\,P} S_n\cdot S_{n'} \notag\\ 
   & + \Bp\,\sum_{\langle n,n'\rangle/\notin\,P} S_n\cdot S_{n'} -\sum_{\langle n,n'\rangle} \Lambda\,S_{nz}^2\label{eq.1} 
 \end{align}
 where $C$ denotes NN spins in a zig-zag chain; $D$ refers to spins in a dimer; $P$ refers to a plane, and the coupling parameters were introduced above. To determine the magnon excitations of the model at low temperatures, we used the Holstein-Primakoff (HP) transformation for spins, in the linear spin-wave approximation. Fourier transforming the Hamiltonian written in terms of the HP-boson operators, we determined the Hamiltonian to be diagonalized in order to determine the magnon spin excitations of
 the CE phase. 
 
 \subsection*{Results and Conclusions}
 To obtain the magnetic excitations of the two-dimensional CE-2D phase, for layered manganites, we diagonalized the Hamiltonian matrix obtained by considering $\Ap=\Bp=0$. As mentioned above, for La$_{0.5}$Sr$_ {1.5}$MnO$_4$ the magnon excitations were measured with INS by Senff \emph{et al.}~\cite{Braden}, who also estimated the magnetic couplings relevant to describe the experimental data in terms of a localized spins model. According to this fit: $\F=\Fp=9.98$ meV, $\A=1.83$ meV, and a magnetic anisotropy of magnitude $\Lambda=0.05$ meV is required. In Fig.~\ref{fig2} we show the lower energy magnon branches we obtained for  \Coml{La}{0.5}{Sr}{1.5} using our model with those parameters, superimposed on the measured data and the fit by Senff \emph{et al.}~\cite{Braden}  
 
 \begin{center}
   \scalebox{0.5}{
     \includegraphics[width=1.1\columnwidth,angle=-90]{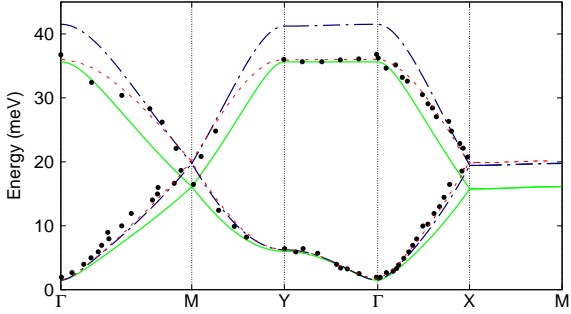}} 
   \captionof{figure}{CE-2D magnons for \Coml{La}{0.5}{Sr}{1.5}. Our calculation (solid line); INS experiment (dotted-line); fit of the exp. data by Senff {\it et al.}~\cite{Braden} (dashed line), and 2D-dimer phase prediction (dot-dashed line). 2D-CE phase parameters: $\F=\Fp=9.98$ meV, $\A=1.83$ meV, $\Lambda=0.05$ meV, $S_1=2\mu_B$ ($\Mn{3}$) and $S_2=1.5\mu_B$ ($\Mn{4}$)~\cite{Braden}. 2D-dimer phase parameters: $\Fp=2\F$, $S_1=S_2=1.75\,\mu_B$, others as for CE-2D.}  
  \label{fig2}
\end{center}

 Our calculations for the CE-2D magnon dispersion relations provide a good qualitative description of the INS data, regarding shape of the curves and the position of the energy maxima and minima. However, at the {\sf M} and {\sf X} Brillouin zone points our CE-2D phase magnon energies differ from the fit of Senff {\it et al.}~\cite{Braden} by about 4.5 meV. This is due to an additional selective FM coupling between second neighbours (2NN) added by Senff et al., who only consider it for 2NN $\Mn{4}$-pairs along a chain (not between 2NN $\Mn{3}$-pairs, which is rather questionable), in order to improve their fit of the measured magnons. 
 
 Senff et al.~\cite{Braden} claim that their results support the presence of the CE phase, and disagree with the dimer phase proposal for half-doped manganites. Nevertheless, Fig.~\ref{fig2} reveals that considering a 2D-dimer model also improves the adjustment at the {\sf M} and {\sf X} BZ points w.r. to the CE-2D phase, without the need of the introduction of a selective second-neighbour FM coupling. Our results, in fact, lend support to the alternative proposal by Efremov et al.~\cite {khomskii} of an "intermediate phase" (between the dimer and CE phases), based on experimental evidence. 
 
 In Fig. 3 we show our prediction for the CE-3D phase magnons in a perovskite manganite, as \Com{La}{0.5}{Sr}{0.5}, estimating the in-plane magnetic coupling parameters as for laminar manganites~\cite{Braden}, and the unknown inter-plane couplings as: $\Bp=1.83$  meV, $\Ap=1.37$ meV, i.e. assuming \mbox{$\Bp\big(\Mn{4}$-$\Mn{4}\big)=\A\big(\Mn{3}$-$\Mn{4}\big)$} and $\Ap<\Bp$ as needed to ensure a stable CE-3D phase. 
 
\begin{center}
  \scalebox{0.5}{
    \includegraphics[width=1.1\columnwidth,angle=-90]{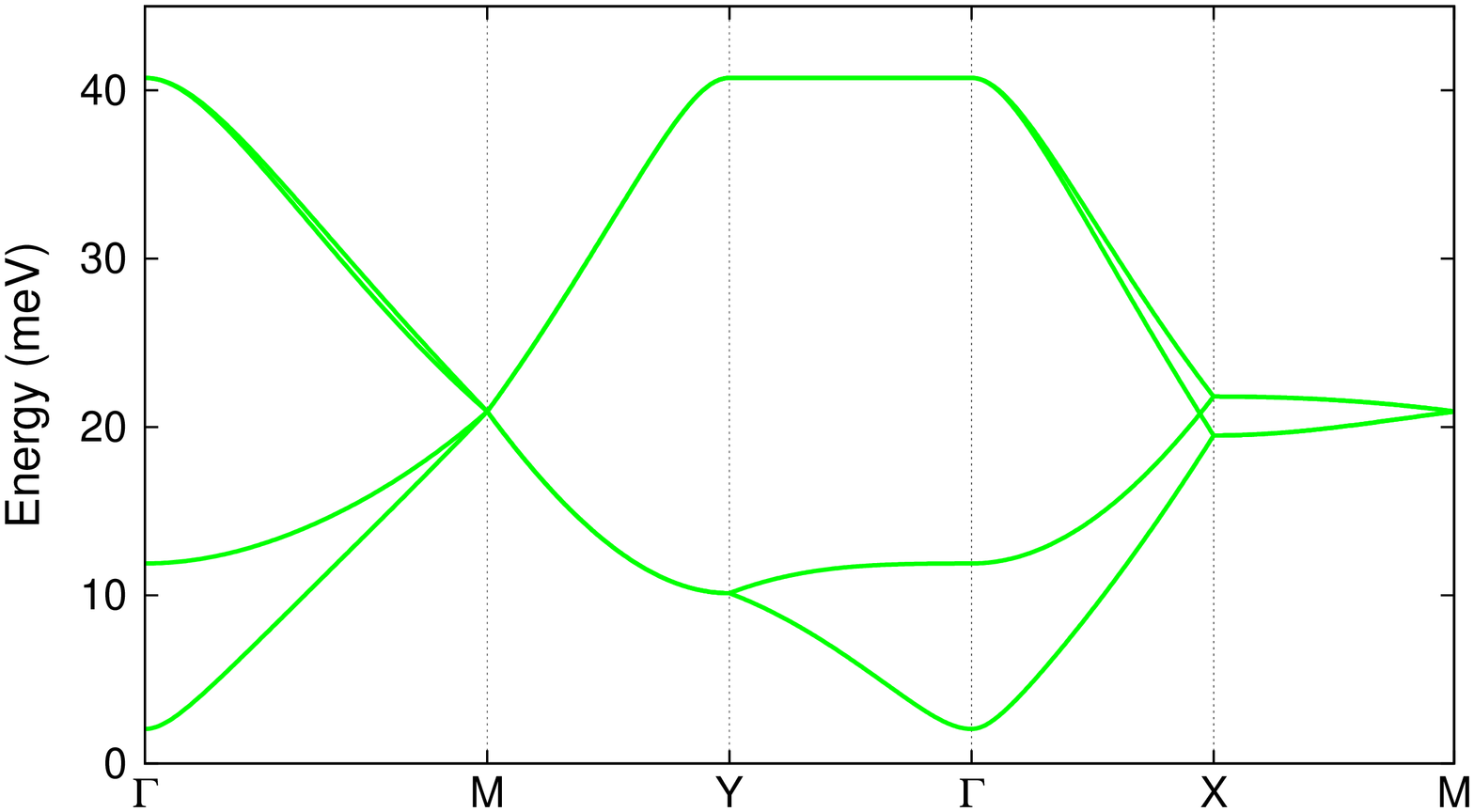}}
  \captionof{figure}{CE-3D low-energy magnons predicted for \Com{La}{0.5}{Sr}{0.5}, with: $\Fp=\F=9.98$ meV; $\A=1.83$ meV; $\Bp=1.83$ meV, $\Ap=1.37$ meV; $c/a_{0}=1$.} 
  \label{fig3}
\end{center} 

 In Fig.4 we exhibit our magnon predictions for the CE-3D phase, the charge-ordered CO-3D and the 3D-dimer phases, evidencing the main differences which should allow to distinguish between them when INS measurements become available. 
\begin{center}
  \scalebox{0.5}{
    \includegraphics[width=1.1\columnwidth,angle=-90]{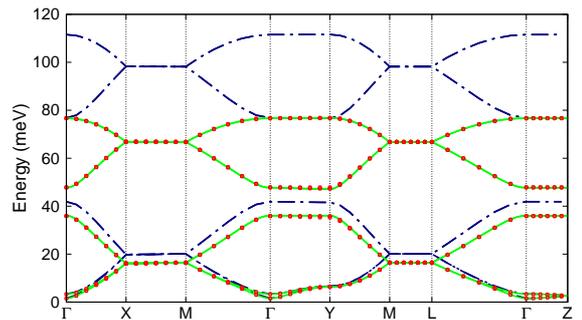}} 
  \captionof{figure}{Magnons predicted for: CE-3D (solid line), CO-3D (dotted- line) and 3D-dimer (dot-dashed line) phases. $\F=9.98$ meV ($=\Fp$: CE+CO; $\Fp= 2\F$: dimer phase); $\A=1.83$ meV; $\Bp=\Ap=0.1$ meV; $c/a_{0}=1$, $\Lambda = 0.05$.} 
  \label{fig4}
\end{center}

\end{multicols}

\end{document}